\documentclass[aps,pra,twocolumn,groupedaddress]{revtex4}
\usepackage{amssymb}
\usepackage{amsmath}
\usepackage{amsthm}
\usepackage{amsfonts}
\usepackage{upgreek}
\usepackage{enumerate}

\usepackage{graphicx}
\usepackage{dcolumn}
\usepackage{bm}

\begin{document}

\title{Influence of the $6^{1}S_{0}$--$6^{3}P_{1}$ Resonance on Continuous Lyman-$\alpha$ Generation in Mercury}
\author{Daniel Kolbe}
\email[]{kolbed@uni-mainz.de}
\author{Martin Scheid}
\altaffiliation{Carl Zeiss Laser Optics GmbH, Carl Zeiss Strasse, D-73446 Oberkochen, Germany}
\author{Jochen Walz}
\affiliation{%
Institut f\"ur Physik, Johannes Gutenberg-Universit\"at Mainz and Helmholtz-Institut Mainz, D-55099 Mainz, Germany
}%

\date{\today}

\begin{abstract}
Continuous coherent radiation in the vacuum-ultraviolet at 122\,nm (Lyman-$\alpha$) can be generated using sum-frequency mixing of three fundamental laser beams in mercury vapour. One of the fundamental beams is at 254\,nm wavelength, which is close to the $6^{1}S_{0}$--$6^{3}P_{1}$ resonance in mercury. Experiments have been performed to investigate the effect of this one-photon resonance on phasematching, absorption and the nonlinear yield. The efficiency of continuous Lyman-$\alpha$ generation has been improved by a factor of 4.5.
\end{abstract}

\maketitle

\section{Introduction}

Coherent tunable radiation in the vacuum-ultraviolet (VUV) can be generated using four-wave mixing (FWM) in metal vapors and gases \cite{Vidal92,Marangos90}. One wavelength of particular interest is $121.56\,\text{nm}$, the Lyman-$\alpha$ transition in atomic hydrogen. Recently the production of cold antihydrogen (a bound state of a positron and an antiproton) \cite{Andresen08,Gabrielse08} and the first trapping \cite{Andresen10,Gabrielse11} in a magnetic trap has been reported. This promises tests of fundamental symmetry between matter and antimatter at ultrahigh precision by comparing the spectra of ordinary hydrogen with its antimatter counterpart \cite{Haensch93} and the first direct test of the equivalence principle for antimatter \cite{Perez05}. A cooling laser on the 1$S$-2$P$ Lyman-$\alpha$ transition will be essential for high precision experiments with antihydrogen. Laser cooling of ordinary hydrogen atoms in a magnetic trap has been demonstrated some time ago using a pulsed Lyman-$\alpha$ source \cite{Setija93} and many pulsed Lyman-$\alpha$ sources have been realized, see e.g.  \cite{Batishe77,Mahon78,Cotter79,Wallenstein80,Cabaret87}. Distinct advantages are expected from laser-cooling using continuous wave (cw) radiation, including reduced spurious pumping into untrapped magnetic substates and higher cooling rates.
 
\begin{figure}[htb]
 \includegraphics[width=0.3\textwidth]{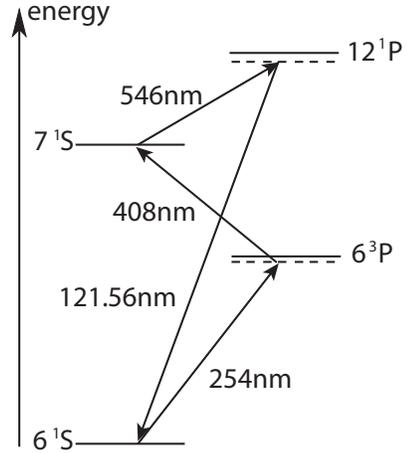}%
 \caption[Level scheme]{Partial energy-level diagram of mercury and the FWM scheme. The UV beam at 254\,nm can be tuned to the $6^1S - 6^3P$ resonance to study the influence on the FWM process. The blue laser at 408\,nm establishes the two-photon resonance with the $7^1S$ state. To generate a beam at Lyman-$\alpha$ (121.56\,nm), the wavelength of the green laser is 546\,nm.}\label{Fig:1}
\end{figure}
 
We already demonstrated a cw Lyman-$\alpha$ source based on solid state lasers \cite{Scheid09}. Figure~\ref{Fig:1} shows the FWM scheme and the corresponding mercury energy levels. Three fundamental beams at $254\,\text{nm}$, $408\,\text{nm}$, and $546\,\text{nm}$ wavelength are used to generate the sum-frequency at Lyman-$\alpha$. The combination of the ultraviolet (UV) beam at $254\,\text{nm}$ and the blue beam at $408\,\text{nm}$ is tuned to the $6^1S - 7^1S$ two-photon resonance, which boosts the Lyman--$\alpha$ yield. Tuning the UV beam close to the $6^1S - 6^3P$ resonance greatly increases the nonlinear susceptibility of the FWM process which can increase the FWM efficiency further \cite{Koudoumas92}. The sum-frequency at Lyman-$\alpha$ is in-between bound states of mercury; the nearest singulett state is the $12^1P$ state. One of the fundamental laser systems can be tuned widely across the $6^1S_0 - 6^3P_0$ resonance in mercury. Hence, this laser system allows investigating the influence of this resonance for the first time. This resonance promises high gain in the nonlinear susceptibility of the FWM process. In this work we present an investigation of near one photon resonant FWM for Lyman-$\alpha$ generation. The paper is organized as follows: First we describe the experimental setup, then we summarize the theoretical background and in Section~IV we present the experimental results.

\section{Experimental setup}

\begin{figure}[htb]
 \includegraphics[width=0.45\textwidth]{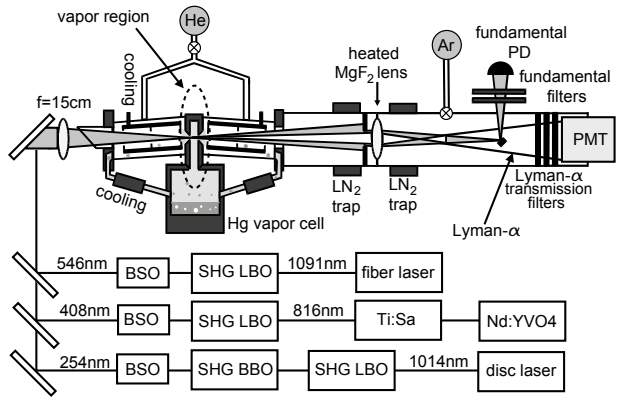}%
 \caption[Experimental setup]{Experimental setup. Lower part: The laser system to generate the fundamental beams (SHG: frequency doubling cavity, BSO: beam shaping optics, LBO, BBO: nonlinear crystals). Upper part: The fundamental beams are focused into the Hg vapor cell for four-wave sum-frequency mixing. Lyman-$\alpha$ radiation is separated from the fundamental beams using a MgF$_2$ lens and is detected with a photomultiplier (PMT). The fundamental beams can be monitored with a photodiode (PD).}\label{Fig:2}
\end{figure}

A schematic of the experimental setup is shown in Fig.~\ref{Fig:2}. The laser system generating the three fundamental beams is shown in the lower part of the figure. Tunable radiation at $254\,\text{nm}$ for near one photon resonant FWM is generated by a frequency-quadrupled Yb:YAG disc laser (ELS, VersaDisk 1030-50). The emission wavelength of the disk laser is tuned with a Lyot filter and an etalon to $1015\,\text{nm}$, which is far of the Yb:YAG gain maximum at $1030\,\text{nm}$ \cite{Spuehler01}. Frequency-quadrupling is done with two subsequent resonant enhancement cavities, the first one using a lithium triborate crystal (LBO) as the nonlinear medium, the second one using a $\beta$-barium borate crystal (BBO). Up to 750\,mW of UV power can be generated from an infrared input power of 4.9\,W. Details of this system have been described elsewhere \cite{Scheid07}. To avoid degradation of the BBO crystal \cite{Kondo98} we typically limit the infrared power to $2\,\text{W}$ which generates $200\,\text{mW}$ of UV radiation. The second fundamental beam at $408\,\text{nm}$ is generated by a frequency-doubled titanium:sapphire laser (Coherent, 899-21), pumped by a frequency doubled Nd:YVO$_4$ laser (Coherent, V10). The external frequency-doubling cavity uses LBO as the nonlinear medium. $1.6\,\text{W}$ of near-infrared radiation at $816\,\text{nm}$ is used to generate a maximum of $500\,\text{mW}$ of blue radiation. The typical day-to-day blue power for our experiments is $300\,\text{mW}$. The third fundamental beam at $546\,\text{nm}$ is generated with a 10\,W fiber laser system at 1091\,nm (Koheras, Adjustik and Boostik) and a modified commercial frequency-doubling cavity (Spectra Physics, Wavetrain). This system is capable of producing up to $4\,\text{W}$ of green radiation \cite{Markert07}. However, at these high powers spontaneous damage of the entrance facet of the amplification fiber has happend two times in two years. Getting the laser repaired by the manufacturer was tedious and very time-consuming. For the present experiments we therefore operate the fiber laser at a very conservative power level of $740\,\text{mW}$ which still gives $280\,\text{mW}$ of green light. 

Beam shaping of each fundamental beam is performed by a pair of cylindrical lenses and spherical telescopes enlarge the beams to allow tighter focusing. The beams are then overlapped with dichroic mirrors and focused into the mercury cell using a fused silica lens (focal length $15\,\text{cm}$). The overlap of the fundamental beams is very critical. For the alignment within the mercury cell the reflections at the entrance window are steered through a $20\,\mu\text{m}$ pinhole. The confocal parameters of the fundamental beams are $b_{254\,\text{nm}}=0.6\,\text{mm}$, $b_{408\,\text{nm}}=0.8\,\text{mm}$ and $b_{546\,\text{nm}}=1.3\,\text{mm}$. A non Gaussian beam profile is caused by the walk-off effect of the nonlinear crystals of the frequency doubling stages and the $M^2$ values are $M^2_{254\,\text{nm}}=1.7$, $M^2_{408\,\text{nm}}=1.6$ and $M^2_{546\,\text{nm}}=1.5$. The mercury cell and the detection region is evacuated to a residual pressure of $10^{-7}\,\text{mbar}$ to prevent absorption of the generated VUV radiation. The FWM takes place in a mercury vapor region of $15\,\text{mm}$ length. The generated Lyman-$\alpha$ beam is separated from the fundamental beams using the dispersion of a MgF$_2$ lens ($f\mathord=21.5\,\text{cm}$ at $546\,\text{nm}$, $f\mathord=13\,\text{cm}$ at Lyman-$\alpha$). A tiny mirror is placed in the focus of the fundamental beams to reflect them out of the vacuum chamber where they can be detected independently. The Lyman-$\alpha$ focus is several centimeters closer to the MgF$_2$ lens and the Lyman-$\alpha$ beam is wide at the position of the tiny mirror. Therefore, the small mirror just casts a shadow in the Lyman-$\alpha$ beam, which causes a loss of about $30\%$. Stray light from the fundamental beams is suppressed further by three VUV interference filters (Acton, 2x122-N and 1x122-XN) and the radiation at Lyman-$\alpha$ is then detected with a solar-blind photomultiplier (Hamamatsu, R6835). 

For small UV detunings to the $6^1S - 6^3P$ resonance we observe an increase of the VUV background. This can be caused due to ionisation from a three photon absorption and subsequent recombination \cite{Bahns97}. The radiative spontaneous decay of the higher levels generates VUV radiation. This background signal is eliminated by chopping the green laser at 1\,Hz and subtracting background from the Lyman-$\alpha$ signal. Great effort has to be made to prevent hot mercury vapor escaping through the beam apertures (diameter 1\,mm) from fogging the entrance window and the separation lens. Additional water cooled apertures outside the vapor region stop the mercury atoms by condensation. In addition both the entrance window and the separation lens are heated to about 60$\,^{\circ}$C to prevent condensation. To further increase the condensation in the cooled region Helium buffer gas can be filled in the vacuum chamber. For mercury vapour temperatures of 220$\,^{\circ}$C a buffer gas pressure of 70-100\,mbar Helium is needed \cite{Eikema01}. Residual gas from the mercury cell is pumped away before every experimental day when the cell is still cold. The pumping time has an effect on the day-to-day reproducibility but on a time-scale of several hours Lyman-$\alpha$ generation is stable.

\section{Theory}\label{theoryIII}

The VUV power generated by sum-frequency four-wave mixing of three Gaussian fundamental beams with confocal parameter $b$ is \cite{Bjorklund75}:

\begin{equation}
 P_{4} =\frac{9}{4} \frac{\omega_1 \omega_2 \omega_3 \omega_4}{\pi^2\epsilon_0^2 c^6}\frac{1}{b^2} \frac{\left|\chi^{(3)}\right|^2}{\Delta k^2} P_1 P_2 P_3 G(b \Delta k) 
\qquad .
\label{Glng:BjorklundG}
\end{equation}

Here $P_4$ is the power at Lyman-$\alpha$ and $P_{1,2,3}$ is the power of the UV, blue and green fundamental beam respectively. The frequencies of the beams are $\omega_{1,2,3,4}$ and $\chi^{(3)}$ is the third-order nonlinear susceptibility. The wavevector mismatch $\Delta k$ is $\Delta k=k_4-k_1-k_2-k_3$ where $k_i$ is the wavevector at frequency $\omega_i$. Both $\Delta k$ in the denominator and $\chi^{(3)}$ in the numerator are proportional to the number density $N$ of mercury atoms. 
The function $G(b \Delta k)$ is called the phase-matching function and contains the dependancy of the Lyman-$\alpha$ power on the number density of mercury atoms. 

We will first discuss the third-order nonlinear susceptibility $\chi^{(3)}$ for two-photon-resonant FWM in mercury, which factorizes \cite{Smith87}

\begin{equation}
\chi^{(3)}=\frac{N}{6 \epsilon_0 \hbar^3} S(\omega_1,\omega_2) \chi_{12} \chi_{34}
\label{chi3}
\qquad ,
\end{equation}

with the two partial susceptibilities 

\begin{equation}
\chi_{12}= \sum_m\left(\frac{ p_{nm} p_{mg} }{\omega_{gm}-\omega_1}+ \frac{p_{nm} p_{mg} }{\omega_{gm}-\omega_2}\right)
\label{chi12}
\qquad ,
\end{equation}

\begin{equation}
\chi_{34}= \sum_\nu \left(\frac{ p_{n\nu} p_{\nu g} }{\omega_{g\nu}-\omega_4}+ \frac{ p_{n\nu} p_{\nu g} }{\omega_{g\nu}+\omega_3}\right)
\label{chi34}
\qquad ,
\end{equation}

and the term describing the enhancement due to the two-photon resonance

\begin{equation}
S(\omega_1+\omega_2)=\frac{1}{\omega_{ng}-(\omega_1+\omega_2)}
\label{Glng:Seinfach}
\qquad .
\end{equation}

Summing over $m$ and $\nu$ in the partial susceptibilities includes all exited states that are linked to the $6^1S$ ground state (index g) and the $7^1S$ state (index n) by dipole transitions. The dipole matrix elements $p_{ab}$ can be obtained from the oscillator strengths $f_{ab}$ tabulated in \cite{Alford87}. For the calculation of $S(\omega_1+\omega_2)$ the broadening of the two-photon resonance has to be taken into account. The homogeneous line-broadening of spontaneous decay and pressure-broadening is included by adding the term $-i\Gamma^{hom}_{7S}/2$ in the denominator of Eq.~(\ref{Glng:Seinfach}). Doppler-broadening is included by adding the Doppler-shift $kv$ of an atom with velocity $v$ and integrating over the one dimensional Boltzmann velocity distribution. One-photon resonances are described by the function $\chi_{12}$ for the UV and blue beam and by $\chi_{34}$ for the green and the resulting Lyman-$\alpha$ beam. In our case with the frequency of the UV beam being very close to the $6^1S_0 - 6^3P_1$ one-photon resonance the second term in $\chi_{12}$ and all the other exited states in the summation except for the $6^3P$ can be neglected. This approximation changes $\chi_{12}$ by 10\% at an UV detuning of 400\,GHz and even less for smaller detunings. In this approximation the nonlinear susceptibility $\chi^{(3)}$ is proportional to $1/(\omega_{6^1S-6^3P}-\omega_1)$, which shows the massive enhancement in the third-order nonlinear susceptibility possible with small UV detuning.

The second factor of interest is the phasematching function $G(b \Delta k)$. This function determines the optimal wavevector mismatch and is maximized for $b \Delta k=-4$ \cite{Bjorklund75}. Phase-matching can be done by adding another gas, such as Kr. This, however, causes pressure-broadening of the mercury resonances, which decreases the Lyman-$\alpha$ yield. We adjust phase-matching by changing the temperature, instead. This changes the mercury vapor density and thus the linear susceptibilities at the fundamental wavelengths and at the sum-frequency. The wavevector mismatch is

\begin{equation}
\Delta k=\frac{1}{c} \left( n_4 \omega_4- n_1 \omega_1- n_2 \omega_2- n_3 \omega_3 \right)
\label{deltak}
\qquad ,
\end{equation} 

where $n_i$ is the index of refraction at frequency $\omega_i$:

\begin{equation}
n_i =1+ \frac{1}{2} {\rm Re}[\chi^{(1)}(\omega_i)]
\label{Glng:n}
\qquad .
\end{equation}

$\chi^{(1)}$ is the first-order susceptibility, which depends on the oscillator strengths $f_{ab}$:

\begin{equation}
\chi^{(1)}(\omega)= N \frac{e^2}{m_e\epsilon_0}\sum_m \frac{f_{gm}}{\omega_{gm}^2-\omega^2}
\label{Glng:chi1}
\qquad .
\end{equation}

with $e$, $m_e$ the electron charge and mass and $\epsilon_0$ the vacuum permittivity. The summation has to be done over all states $m$ connected to the ground state $g$ via a dipole transition at the frequency $\omega_{gm}$. The index of refraction of the UV beam is much higher than for the other beams because the green, blue, and Lyman-$\alpha$ frequencies are far from any one-photon resonance. The resulting wavevector mismatch $\Delta k$ thus is dominated by the near $6^1S_0 - 6^3P_1$ one-photon resonance of the UV beam. The leading term in $\Delta k$ is proportional to $N/(\omega_{6^1S-6^3P}^2-\omega_1^2)$. Thus the phase-matching density and temperature strongly depends of the UV detuning. At the phase matching temperature the wavevector mismatch is $\Delta k=-4/b$, therefore the phase-matching density and temperature becomes lower for smaller UV detuning. This counteracts the enhancement of the third order nonlinear susceptibility: As can be seen in Eq.~(\ref{chi3}) a reduced density results in a smaller nonlinear susceptibility which cancels the enhancement of $\chi_{12}$ by the smaller UV detuning. Nevertheless it is of interest to investigate the influence of a one-photon resonance since the effects of absorption and buffer gas are not included so far.

\section{Measurements}

\subsection{Phase-matching}\label{phase-matching}

\begin{figure}[tb]
 \includegraphics[width=0.35\textwidth,angle=-90]{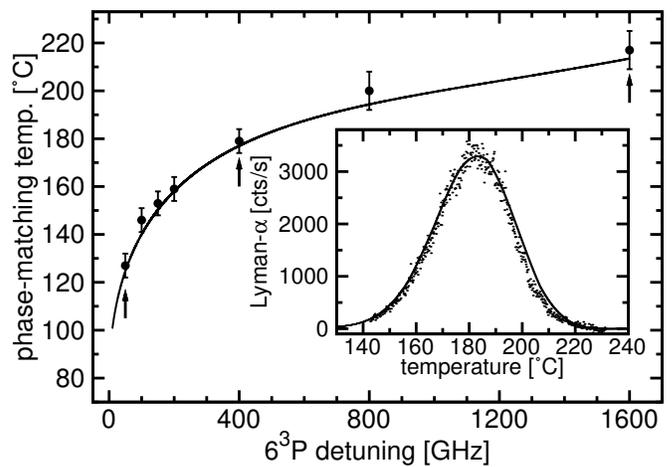}%
 \caption[phase-matching]{Phase-matching temperatures for different UV detunings to the $6^{1}S$--$6^{3}P$ transition. For each UV detuning a phase-matching curve is measured and the temperature with maximum Lyman-$\alpha$ power is determined. The line is a calculation. Inset: Example for a phase-matching curve at a UV detuning of 400\,GHz. Points are measured Lyman-$\alpha$ yield as a function of the cell temperature. The line represents a calculation for which the temperature of the mercury was taken to be $6^\circ$C colder than the measured temperature on the outside of the stainless-steel pipes.}\label{Fig:3}
\end{figure}

Phase-matching of the FWM process is performed by adjusting the temperature and thereby the density of the nonlinear medium. To determine the phase-matching temperature we performed temperature scans of the cell and measured the Lyman-$\alpha$ yield. One example is shown for one specific UV detuning (400\,GHz) in the inset of Fig.~\ref{Fig:3}. The maxima of such phasematching curves are drawn in the main graph of Fig.~\ref{Fig:3} and show the dependency of the phase-matching temperature as a function of the UV detuning. At some detunings more than one phase-matching curve has been measured with deviations of the phase-matching temperature of a few $^{\circ}$C. For those detunings the mean value has been taken (points marked with arrow). The error bars correspond to a drop of the Lyman-$\alpha$ yield of 5\%. As expected from the theoretical considerations the phase-matching temperature decreases with smaller UV detunings. The solid line in the main graph is a calculation. For Gaussian beams with identical confocal parameters the phase-matching temperature can be calculated with Eq.~(\ref{Glng:chi1}) and the phase-matching condition $\Delta k=-4/b$. In this simple case a theoretical phase-matching curve can be calculated with the phase-matching function $G(b\Delta k)$. Different $b$-parameters for Gaussian fundamental beams can be treated using the integral formalism introduced by Lago et al. \cite{Lago87}. This formalism can also incorporate absorption of the involved beams which becomes important for the results in the next section. The theory curves in Fig.~\ref{Fig:3} are calculated with this modified phase-matching function $G(b_{254\,\text{nm}},b_{408\,\text{nm}},b_{546\,\text{nm}}, \Delta k)$. The theoretical phasematching curve in the inset of Fig.~\ref{Fig:3} has been shifted by 6\,$^{\circ}$C to higher temperatures to match the experimental data. This indicates that the mercury vapor and liquid is actually colder than the temperature on the outside of the stainless-steel pipes, an effect which has been observed before \cite{Smith86}. For the calculation we have taken the measured confocal parameters to be equal to the confocal parameters of Gaussian beams. The propagation of a non-Gaussian beam can be described with an embedded pure Gaussian beam with the same confocal parameter. The embedded Gaussian beams give the main contribution to the generated Lyman-$\alpha$ radiation \cite{Kolbe10}. This assumption gives good agreement with the measured data.

In our previous work \cite{Scheid09} we observed a surprisingly large discrepancy of 40\,$^{\circ}$C between the theoretically calculated and the measured phase-matching temperature. Two reasons were found for this mismatch: First, the temperature scan were now done with a smaller temperature slope ($0.008^\circ$C/s instead of $0.08^\circ$C/s), so that the mercury vapor has more time for thermalization. This is very crucial since the heat flow from the cell to the mercury is low. With the smaller rate of temperature change the phasematching curves of upward and downward temperature scans are equal. Compared to our previous results this shifts the experimental phase-matching curve by 17\,$^{\circ}$C to lower temperatures. Second, the calculation is done with the three measured b-parameters of the fundamental beams rather then using one average $b$-parameter for all three beams. This shifts the theoretical curve by 16\,$^{\circ}$C to higher temperatures.

\subsection{Absorption}\label{absorption}

\begin{figure}[tb]
 \includegraphics[width=0.35\textwidth,angle=-90]{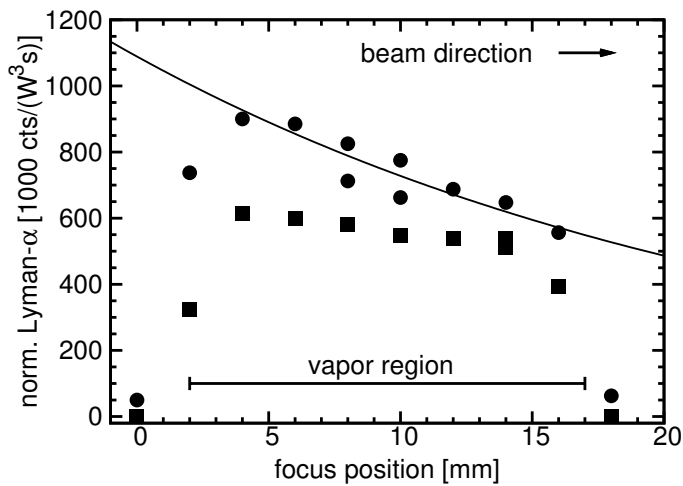}%
 \caption{Lyman-$\alpha$ yield as a function of the focus position in the mercury cell. The extension of the mercury vapor zone is from 2 to 17\,mm outside the mercury vapour zone no Lyman-$\alpha$ is generated. Circles are measured at a UV detuning of 50\,GHz (130$^\circ$C), squares at 1600\,GHz (220$^\circ$C). For small detunings with a higher UV absorption a focus position near the beginning of the vapor region is preferable. The solid line is an exponential fit to the data points at 50\,GHz detuning in the region from 4 to 16\,mm.} 
\label{Fig:4}
\end{figure}

The $6^1S - 6^3P$ transition causes absorption at the UV wavelength which reduces the fundamental UV power. Due to absorption along the propagation through the mercury cell less UV power is available for the FWM at the focus position. The effect of the UV absorption is shown in Fig.~\ref{Fig:4}. In this measurement we have changed the position of the fundamental foci by changing the position of the focusing lens. Squares are measured at a UV detuning of 1600\,GHz, where no absorption of UV light through the whole apparatus was observed. The circles are measured at an UV detuning of 50\,GHz, here we measured 52\% of absorption of the UV light through the cell. The Lyman-$\alpha$ yield is normalized on the fundamental powers to remove the effect of drifts of the intensity of the fundamental beams on the measurement. As can be seen from the 1600\,GHz data points the density profile in the mercury vapor cell is approximately rectangular with a steep edge at the ends of the cell. Within the vapor region only a small effect of the focus position can be distinguished. The increased absorption at 50\,GHz UV detuning causes a larger influence of the position of the focus on the generated Lyman-$\alpha$ radiation. The absorption of the UV radiation can be included in the modified phase-matching of the integral-formalism. With the parameters of our experiment the effect of UV absorption before the focus clearly dominates over the effect of absorption in the focus region itself. Then one can use the much simpler Eq.~(\ref{Glng:BjorklundG}) with the remaining part of the UV power for $P_1$. Assuming a rectangular density profile for the mercury vapour one then gets an exponential decay with the absorption coefficient as the damping constant. The solid line in Fig.~\ref{Fig:4} is an exponential fit to the data points at 50\,GHz detuning in the vapor region. This gives an absorption coefficient of $\alpha=0.04$\,mm$^{-1}$. Assuming a 15\,mm long rectangular density profile this results in an absorption of 55\% over the cell, which agrees with the measured UV absorption of 52\% through the whole apparatus.

\subsection{Two-photon resonance and buffer gas}\label{buffer gas}

\begin{figure}[tb]
\includegraphics[width=0.35\textwidth,angle=-90]{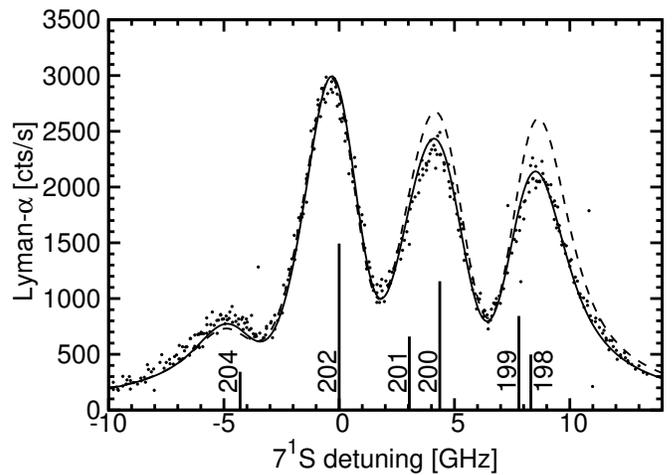}%
\caption{Lyman-$\alpha$ yield as a function of the detuning of the blue laser, which is scanned over the $6^1S_0 - 7^1S_0$ two-photon resonance. The UV detuning is 50\,GHz. The dashed line is calculated neglecting of the $6^1S_0 - 6^3P_1$ resonance splitting. The solid line is calculated including the one-photon resonance structure. Vertical bars at the bottom indicate the relative abundances of the different isotopes and their isotopic shift.} 
\label{Fig:5}
\end{figure}

The influence of the $6^1S - 7^1S$ two-photon resonance on the FWM in mercury vapor has been thoroughly investigated \cite{Smith87,Scheid09}. In these experiments the UV radiation was far detuned from the $6^1S - 6^3P$ one-photon resonance. The line shape of the two-photon resonance can then be described by the function $S(\omega_1+\omega_2)$ of Eq. \ref{Glng:Seinfach} with homogeneous and Doppler-broadening added. Figure~\ref{Fig:5} shows the Lyman-$\alpha$ yield as a function of the blue laser detuning. The different maxima arise from the individual mercury isotopes and the isotope splitting of the $6^1S - 7^1S$ two-photon resonance \cite{Gerstenkorn77}. The dashed line in Fig.~\ref{Fig:5} is calculated with the function $S(\omega_1+\omega_2)$ and does not fit the experimental data adequately. Contributions of the isotopes $^{201}$Hg-$^{198}$Hg with larger UV detuning are overestimated. The UV detuning in this measurement was only 50\,GHz with respect to the most abundant $^{202}$Hg isotope. For such small detunings the isotope shifts of the $6^1S - 6^3P$ transition, which is in the order of 1-10\,GHz, become important as well. The UV detuning differs for the individual mercury isotopes. Thus the two-photon resonance of the isotopes with smaller UV detuning is enhanced and the two-photon resonance for isotopes with larger UV detuning is lowered. The solid curve is calculated with the function $\chi_{12}S(\omega_1+\omega_2)$ but including the isotope splitting of the $6^1S - 6^3P$ resonance (obtained from \cite{Schweitzer63}) included in $\chi_{12}$. This function is in good agreement with the experimental data, which demonstrates the importance of the $6^1S - 6^3P$ isotope splitting for small UV detunings.

\begin{figure}[tb]
 \includegraphics[width=0.35\textwidth,angle=-90]{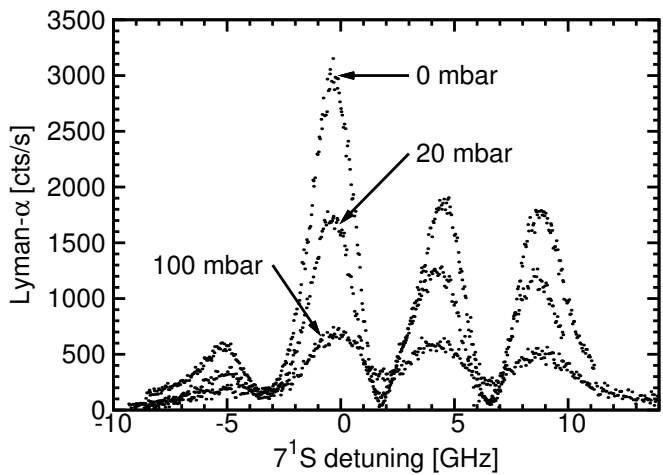}%
 \caption{Lyman-$\alpha$ yield as a function of the blue laser detuning at different helium buffer gas pressures. Lower helium pressures causes less pressure broadening and thus narrows and heightens the two-photon resonance. Hence, the Lyman-$\alpha$ yield is enhanced for lower helium pressures.} 
\label{Fig:6}
\end{figure}

For decreased UV detuning the phase-matching temperature of the FWM process becomes smaller as well, see Fig. \ref{Fig:3}. At lower temperatures there is less flux of mercury atoms trough the beam access apertures. Thus less helium buffer gas pressure is needed to prevent mercury condensation onto the optics. This has two positive effects: The foreign pressure broadening of the $6^1S - 6^3P$ transition due to the buffer gas becomes smaller which reduces the UV absorption. Also the $6^1S - 7^1S$ two-photon resonance becomes both narrower and higher. Thus with lower buffer gas pressures more Lyman-$\alpha$ can be generated. This is shown in Fig.~\ref{Fig:6} where the two-photon resonance of the FWM process has been measured for different buffer gas pressures and fixed UV detuning of 50\,GHz. At 50\,GHz the phase-matching temperature is only 130$\,^{\circ}$C. At such a low temperature the mercury cell can be operated even without buffer gas. For the measurement at 50\,GHz this gives a factor of 4.3 more Lyman-$\alpha$ power compared to a buffer gas pressure of 100\,mbar, typically used for the higher UV detunings. In Fig.~\ref{Fig:6} at 0\,mbar buffer gas pressure the combined peaks at 4\,GHz ($^{201}$Hg and $^{200}$Hg) and 8.5\,GHz ($^{199}$Hg and $^{198}$Hg) are slightly shifted towards larger detunings. This effect is not reproduced using the theory of Section~\ref{theoryIII} and would need further investigation.

\subsection{One-Photon Resonance}\label{One-Photon-Resonance}

\begin{figure}[tb]
 \includegraphics[width=0.35\textwidth,angle=-90]{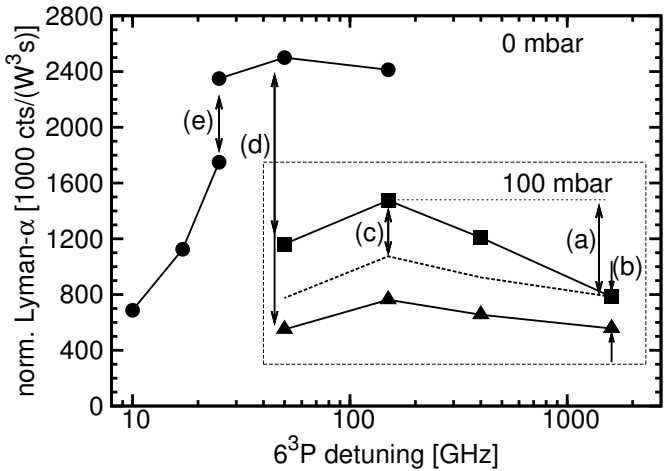}%
 \caption{Lyman-$\alpha$ enhancement due to the $6^{1}S$--$6^{3}P$ resonance. The triangles are measured with focus position in the middle of the vapor region, the squares with the focus at the beginning (both at a helium pressure of 100\,mbar). The circles are measured without helium buffer gas and focus at the beginning of the vapor region. The lines connect datasets recorded at the same day.  The arrows are discussed in the text.} 
\label{Fig:7}
\end{figure}

Figure~\ref{Fig:7} summarizes the results of our experiments on the effect of the near $6^{1}S$--$6^{3}P$ one-photon resonance on the Lyman-$\alpha$ yield. Three different situations have been tested: First with buffer gas inside the mercury cell with focus position in the middle of the cell (triangles), second with the focus position at the beginning of the cell with buffer gas (boxes) and third without buffer gas and the focus position at the beginning of the cell (circles). Data points which are connected by lines were measured at the same day.

Let us first discuss the dataset at 100\,mbar buffer gas pressure with the focus at the beginning of the cell (boxes). A small positive effect of smaller detuning with respect to the $6^{1}S$--$6^{3}P$ one-photon resonance can be seen. The overall effect is about a factor of 1.9 indicated by arrow (a). The maximum efficiency is at 150\,GHz and at smaller detunings absorption at the UV wavelength reduces the efficiency. The enhancement of the Lyman-$\alpha$ power at smaller UV detuning to the $6^{1}S$--$6^{3}P$ one-photon resonance is not as high as might naively be expected from the resonance denominator in the nonlinear susceptibility (see Eq.~(\ref{chi12})). The lower mercury density at lower phasmatching temperatures counteracts and cancels the resonant enhancement. From the theory some small enhancement (of about 7\%) is expected because of the reduced Doppler-broadening at lower phase-matching temperatures. The measured enhancement is somewhat higher.

Let us now compare the datasets in the dashed region of Fig.~\ref{Fig:7} rectangular which are both measured at 100\,mbar buffer gas pressure but at different focus positions within the cell. The dataset with triangles was measured with the focus in the middle of the mercury cell. The FWM efficiency is increased by choosing a focus position closer to the entrance of the cell where less UV power is absorbed before reaching the focus. Since at 1600\,GHz detuning only low influence of the focus position on the four-wave mixing should be expected (see Fig.~\ref{Fig:4}), we assume that the observed offset of the two datapoints at 1600\,GHz (indicated by arrow (b)) is caused by day-to-day reproducibility. To better compare these two datasets we scaled the lower dataset so that the datapoints at 1600\,GHz are the same (dashed line). The influence of the focus position is larger for smaller detunings. At the maximum the Lyman-$\alpha$ efficiency increases by about 30\%, indicated by arrow (c) in Fig.~\ref{Fig:7}.

At small detunings ($<$150\,GHz) at lower phasematching temperatures the buffer gas can be omitted which results in the dataset with circles. The reduced pressure broadening influences four-wave mixing in two ways: First, the increase of the two-photon resonance (see Fig.~\ref{Fig:6}) improves the four-wave mixing efficiency. Second the reduced absorption of UV light caused by the smaller one-photon linewidth shifts the maximum to lower detunings (50\,GHz) and even at a detuning of 10\,GHz four-wave mixing can be still observed. The enhancement is between a factor of 2.2 and 4.5 compared to the datasets with buffer gas at a detuning of 50\,GHz indicated by arrow (d). The two datasets at 0\,mbar buffer gas pressure were recorded at different days and the arrow (e) at the data at 25\,GHz detuning again shows the day-to-day reproducibility.

The highest count rate normalized to the powers in the cell is $2.5\times10^6$\,cts/(W$^3$ s) at 50 GHz without buffer gas and the focus at the beginning of the vapor region. This is an improvement of a factor 3 compared to our previous results \cite{Scheid09} with a UV detuning of 400\,GHz. The highest measured absolute count rate is 10800\,cts/s with fundamental powers of $P_{1}=130$\,mW, $P_{2}=324$\,mW and $P_{3}=185$\,mW in the mercury cell which corresponds to a conversion efficiency of $1.1\times10^6$ cts/(W$^3$ s). The overall detection efficiency of Lyman-$\alpha$ due to the MgF$_2$ lens, the small mirror, the filters, and the photomultiplier efficiency is $6\times10^{-5}$. Therefore the Lyman-$\alpha$ power generated is 0.3\,nW. This is actually somewhat lower than the 0.4\,nW published earlier \cite{Scheid09} because we had underestimated the detection efficiency. For the present work, we have measured the transmission of the MgF$_2$ lens and the filters in a separate apparatus using a deuterium lamp and a VUV monochromator. With full fundamental powers of $P_{1}=750$\,mW, $P_{2}=500$\,mW and $P_{3}=4$\,W the system is capable of generating a Lyman-$\alpha$ power of up to 140\,nW. This is sufficient for future laser cooling of anti-hydrogen on a time-scale of minutes \cite{Walz01}.

\section{Conclusions}

Continuous coherent radiation at Lyman-$\alpha$ can be generated by four-wave mixing in mercury vapor. Tuning one fundamental wavelength close to a one-photon resonance in mercury influences several quantities such as the nonlinear susceptibility, the phase-matching temperature, and absorption. In this paper a comprehensive study of these effects in continuous four-wave sum-frequency mixing has been presented for the first time. One outcome is an overall efficiency enhancement of a factor of 4.5 by the use of the $6^{1}S$--$6^{3}P$ resonance in mercury. This is a significant improvement of the Lyman-$\alpha$ yield for future laser-cooling of antihydrogen.

\begin{acknowledgments}
This work was supported by the German Ministry of Education and Research (BMBF). 
\end{acknowledgments}
\bibliography{Kolbe_bib}

\end{document}